\documentstyle[aps,eqsecnum,preprint]{revtex}
\tightenlines
\begin{document}
\draft
\preprint{}
\title{Isoscalar resonances with $J^{PC}=1^{--}$
in $e^+e^-$ annihilation.}
\author{N.~N.~Achasov \footnote{e-mail: achasov@math.nsc.ru}
 and A.~A.~Kozhevnikov \footnote{e-mail: kozhev@math.nsc.ru}}
\address{Laboratory of Theoretical Physics,\\
S.L.~Sobolev Institute for Mathematics,\\
630090, Novosibirsk 90, Russian Federation}
\date{\today}
\maketitle
\language=1
\begin{abstract}
The analysis of the vector isoscalar excitations in the energy range
between 1 and 2 GeV of the $e^+e^-$ annihilation is presented for the final
states $\pi^+\pi^-\pi^0$, $\omega\pi^+\pi^-$, $K^+K^-$, $K^0_SK^\pm\pi^\mp$
and $K^{\ast0}K^-\pi^++\mbox{c.c}$. The effects
of both  resonance mixing and the successive opening of multiparticle
channels, with  energy-dependent partial widths, are taken into account.
The work extends
our previous analysis of vector isovector excitations and is
aimed at comparing the existing data with the predictions of the $q\bar q$
model. It is shown that this hypothesis does not contradict the data.
\end{abstract}


\narrowtext

\section{Introduction}
\label{intro}

The issue of  excitations with quantum numbers $J^{PC}=1^{--}$
\cite{pdg} in the
energy range between 1 and 2 GeV of  $e^+e^-$ annihilation still
remains, to a large extent, an unresolved one. Are they recurrences of
the ground state nonet of $\rho(770)$, $\omega(782)$, and $\varphi(1020)$,
or do they have an exotic nature
\cite{bityuk87,lipkin87,ach86,ach88,kalash,clegg94,page95}?
In the former case, to
what extent are the flavor SU(3) predictions good for them? In the latter
case, are they completely exotic, and if not, what is the admixture of the
exotic non-$q\bar q$ component? To answer these and similar questions, one
should extract the masses and coupling constants of bare resonances in order
to
compare them with various models. As we have shown earlier \cite{ach97a}
in the case of the vector isovector $\rho$-like excitations, taking into
account both the effects of resonance mixing and the fast energy growth
of the partial widths of successively opened multiparticle channels affects
the specific masses and coupling constants extracted from the data.

The present paper is aimed at extending a similar treatment to the case of
vector isoscalar  $\omega-$ and $\varphi-$like excitations. To this
end we analyze the data on the reactions
\begin{equation}
e^+e^-\to\pi^+\pi^-\pi^0
\label{dm2_3pi}
\end{equation}
\cite{nd,dm2_3pi,cmd},
\begin{equation}
e^+e^-\to\omega\pi^+\pi^-
\label{om2pi}
\end{equation}
\cite{dm2_3pi},
\begin{equation}
e^+e^-\to K^+K^-
\label{dm2_kk}
\end{equation}
\cite{olya,dm2_kk},
\begin{equation}
e^+e^-\to K^{\ast0}\bar K^0+\mbox{c.c}\to K^0_SK^\pm\pi^\mp
\label{kskpi}
\end{equation}
\cite{dm2_kstk},
\begin{equation}
e^+e^-\to K^{\ast0}K^-\pi^+(\bar K^{\ast0}K^+\pi^-)
\label{dm2_kstkpi}
\end{equation}
\cite{dm2_kstkpi}, allowing for  contributions of the
$\omega^\prime_{1,2}$ $\varphi^\prime_{1,2}$ resonances, in order
to extract the masses and coupling constants
of these excitations to various channels.
The main result is that, within very large errors
determined by poor data samples, the extracted parameters do not contradict
the simple $q\bar q$ model of the $\omega$- and $\varphi$-like resonances.
The magnitudes of the $q\bar q$ bound state wave function at
the origin are extracted from the magnitudes of the leptonic widths,
yet the accuracy
still does not permit one to verify the traditional $q\bar q$ assignment
\cite{pdg,godfrey85}  of  heavier excitations and to draw any definite
conclusions about the interquark potential.

The paper is organized as follows. Section \ref{sec2} contains the
expressions for the cross sections and a discussion of the assumptions
made about the interaction vertices and the coupling constants.
Section \ref{sec3} is 
devoted to the presentation of the results of our analysis,
which are discussed in Sec. \ref{sec4}. Section \ref{sec5} sketches possible
further work necessary for the improvement of the situation with 
excitations with  masses above 1 GeV.

\section{Basic formulas required for the analysis}

\label{sec2}

\subsection{Expressions for the cross sections}
\label{susec2a}
An exact application of the explicitly unitary method \cite{ach84} of
taking the mixing of resonances into account is computationally
time consuming in the present case,
since it demands the inversion of the $9\times9$
matrix of inverse propagators whose  elements are complex numbers.
Instead, we take into account
the mixing inside each sector, $\rho(770)-\rho^\prime_1
-\rho^\prime_2$, $\omega(782)-\omega^\prime_1
-\omega^\prime_2$ or $\varphi(1020)-\varphi^\prime_1
-\varphi^\prime_2$ to all orders, while the terms which break the
Okubo-Zweig-Iizuka (OZI) rule are taken into account to first order. Then
the cross section of production of the final state $f$ in $e^+e^-$
annihilation can be represented as
\begin{eqnarray}
\sigma_f&=&
{(4\pi\alpha)^2\over s^{3/2}}|\left(g_{\gamma\rho},
g_{\gamma\rho^{\prime}_1},g_{\gamma\rho^{\prime}_2}\right)G^{-1}_\rho(s)
\left(
\begin{array}{c}
g_{\rho f}\\ g_{\rho^{\prime}_1 f}\\ g_{\rho^{\prime}_2 f}
\end{array}
\right)\;
                    \nonumber\\
& &+\left(g_{\gamma\omega},
g_{\gamma\omega^{\prime}_1},g_{\gamma\omega^{\prime}_2}\right)
G^{-1}_\omega(s)
\left(
\begin{array}{c}
g_{\omega f}\\ g_{\omega^{\prime}_1 f}\\ g_{\omega^{\prime}_2 f}
\end{array}
\right)\;
                    \nonumber\\
& &+\left(g_{\gamma\varphi},
g_{\gamma\varphi^{\prime}_1},g_{\gamma\varphi^{\prime}_2}\right)
G^{-1}_\varphi(s)
\left(
\begin{array}{c}
g_{\varphi f}\\ g_{\varphi^{\prime}_1 f}\\ g_{\varphi^{\prime}_2 f}
\end{array}
\right)\;           \nonumber\\
& &+\left(g_{\gamma\omega},
g_{\gamma\omega^{\prime}_1},g_{\gamma\omega^{\prime}_2}\right)
G^{(1)}_{\rm OZI}(s)\left(
\begin{array}{c}
g_{\varphi f}/D_\varphi\\
g_{\varphi^{\prime}_1 f}/D_{\varphi^\prime_1}
\\ g_{\varphi^{\prime}_2 f}/D_{\varphi^\prime_2}
\end{array}
\right)\;           \nonumber\\
& &+\left(g_{\gamma\varphi},
g_{\gamma\varphi^{\prime}_1},g_{\gamma\varphi^{\prime}_2}\right)
G^{(2)}_{\rm OZI}(s)\left(
\begin{array}{c}
g_{\omega f}/D_\omega\\
g_{\omega^{\prime}_1 f}/D_{\omega^\prime_1}\\
g_{\omega^{\prime}_2 f}/D_{\omega^\prime_2}
\end{array}
\right)\;|^2P_f,
\label{eq2.1}
\end{eqnarray}
where $f=\pi^+\pi^-\pi^0$, $\omega\pi^+\pi^-$,  $K^+K^-$, $K^0_SK^+\pi^-$,
$K^{\ast 0}K^-\pi^+$; $s$ is the total
center-of-mass energy squared, $\alpha=1/137$.
The leptonic widths on the mass shell of the unmixed states are expressed
through the $\gamma\to V$ transition amplitudes $g_{\gamma V}$ and the
leptonic coupling constants $f_V$ as usual:
\begin{eqnarray}
\Gamma_{Ve^+e^-}&=&\frac{4\pi\alpha^2g^2_{\gamma V}}{3m^3_V},
                 \nonumber\\
g_{\gamma V}&=&\frac{m^2_V}{f_V}.
\label{eq2.4}
\end{eqnarray}
The matrices entering into Eq. (\ref{eq2.1}) are, respectively,
\begin{equation}
G_V(s)=\left(
\begin{array}{ccc}D_V&-\Pi_{VV^{\prime}_1}&-\Pi_{VV^{\prime}_2}  \\
-\Pi_{VV^{\prime}_1}&D_{V^{\prime}_1}&
-\Pi_{V^{\prime}_1V^{\prime}_2}\\
-\Pi_{VV^{\prime}_2}&-\Pi_{V^{\prime}_1V^{\prime}_2}&
D_{V^{\prime}_2}
\end{array}
\right)\;
\label{eq2.2}
\end{equation}
($V=\rho,\omega,\varphi$) and
\begin{mathletters}
\label{eq2.3}
\begin{equation}
G^{(1)}_{\rm OZI}(s)=\left(
\begin{array}{ccc}
{\Pi_{\varphi\omega}\over D_\omega}&
{\Pi_{\varphi^\prime_1\omega}\over D_\omega}&
{\Pi_{\varphi^\prime_2\omega}\over D_\omega}\\
{\Pi_{\varphi\omega^\prime_1}\over D_{\omega^\prime_1}}&
{\Pi_{\varphi^\prime_1\omega^\prime_1}\over D_{\omega^\prime_1}}&
{\Pi_{\varphi^\prime_2\omega^\prime_1}\over D_{\omega^\prime_1}}\\
{\Pi_{\varphi\omega^\prime_2}\over D_{\omega^\prime_2}}&
{\Pi_{\varphi^\prime_1\omega^\prime_2}\over D_{\omega^\prime_2}}&
{\Pi_{\varphi^\prime_2\omega^\prime_2}\over D_{\omega^\prime_2}}
\end{array}\right)\;,
\end{equation}
\begin{equation}
G^{(2)}_{\rm OZI}(s)=\left(
\begin{array}{ccc}
{\Pi_{\varphi\omega}\over D_\varphi}&
{\Pi_{\varphi\omega^\prime_1}\over D_\varphi}&
{\Pi_{\varphi\omega^\prime_2}\over D_\varphi}\\
{\Pi_{\varphi^\prime_1\omega}\over D_{\varphi^\prime_1}}&
{\Pi_{\varphi^\prime_1\omega^\prime_1}\over D_{\varphi^\prime_1}}&
{\Pi_{\varphi^\prime_1\omega^\prime_2}\over D_{\varphi^\prime_1}}\\
{\Pi_{\varphi^\prime_2\omega}\over D_{\varphi^\prime_2}}&
{\Pi_{\varphi^\prime_2\omega^\prime_1}\over D_{\varphi^\prime_2}}&
{\Pi_{\varphi^\prime_2\omega^\prime_2}\over D_{\varphi^\prime_2}}
\end{array}\right)\;,
\end{equation}
\end{mathletters}
where the inverse propagators of the bare states, $D_V\equiv D_V(s)$,
and the nondiagonal polarization operators
$\Pi_{VV^\prime}\equiv\Pi_{VV^\prime}(s)$ responsible for the mixing are
discussed below in Sec. \ref{susec2c}.
In what follows we will often use also the notation
$V_i$ ($i=1,2,3$) such that $V_1,V_2,V_3$
corresponds to $V,V^\prime_1,V^\prime_2$, and $V=\rho, \omega, \varphi$.

The factor $P_f$ for the  final states $f=\pi^+\pi^-\pi^0$,
$\omega\pi^+\pi^-$,  $K^+K^-$, $K^0_SK^+\pi^-$,
$K^{\ast 0}K^-\pi^+$ reads, respectively,
\begin{eqnarray}
P_f&\equiv& P_f(s)=
\frac{W_{3\pi}(\sqrt{s})}{4\pi}\mbox{, }
W_{VP\pi}(\sqrt{s},m_\omega,m_\pi)\mbox{, }   \nonumber\\
& &\frac{q^3_{KK}}{6\pi s}\mbox{, }
{1\over2}\frac{q^3_{K^\ast\bar K}}{12\pi}
\mbox{, }W_{VP\pi}(\sqrt{s},m_{K^\ast},m_K),
\label{eq2.5}
\end{eqnarray}
where
\begin{eqnarray}
\langle q^3_{K^\ast\bar K}\rangle&=&\int_{(m_K+m_\pi)^2}
^{(\sqrt{s}-m_K)^2}{dm^2m_{K^\ast}\Gamma_{K^\ast}/\pi
\over(m^2-m^2_{K^\ast})^2
+m^2_{K^\ast}\Gamma^2_{K^\ast}}       \nonumber\\
& &\times q^3(\sqrt{s},m,m_{K^\ast})
\label{smear}
\end{eqnarray}
stands for the smearing  implied by the finite width of the $K^\ast$
meson. Hereafter,
\begin{eqnarray}
q_{ij}&\equiv& q(M,m_i,m_j)={1\over2M}\{[M^2-(m_i-m_j)^2] \nonumber\\
& &\times[M^2-(m_i+m_j)^2]\}^{1/2}
\label{eq6}
\end{eqnarray}
is the magnitude of the momentum of either
particle $i$ or $j$, in the rest
frame of the decaying particle.
The origin of the multiplier  1/2 in the case of  $f=K^0_SK^+\pi^-$
is explained below. See Sec. \ref{susec2b}.
Assuming  pointlike dynamics for the vertex
vector$\to VP\pi$, where $V$ stands for the vector
meson, and $P=K,\pi$, the factor of the $VP\pi$ final state can be
written as
\begin{eqnarray}
W_{VP\pi}(\sqrt{s},m_V,m_P)&=&
\frac{1}{(2\pi)^34s}
\int_{m_P+m_\pi}^{\sqrt{s}-m_V}dm     \nonumber\\
& &\times\left(1+\frac{q^2(\sqrt{s},m_V,m)}{3m^2_V}\right)
\nonumber\\
& &\times q(\sqrt{s},m_V,m)       \nonumber\\
& &\times q(m,m_P,m_\pi).
\label{eq2.6}
\end{eqnarray}
The factor
\begin{eqnarray}
W_{3\pi}(\sqrt{s})&=&{g^2_{\rho\pi\pi}\over12\pi^2}
\int_{2m\pi}^{\sqrt{s}-m_\pi}dm q^3_\rho(m) q^3_\pi(m)
\nonumber\\
& &\times\int_{-1}^1dx(1-x^2)\left|{1\over D_\rho(m^2)}\right.
\nonumber\\
& &+\left.
{1\over D_\rho(m^2_-)}+{1\over D_\rho(m^2_+)}\right|^2,
\label{wdm2_3pi}
\end{eqnarray}
where
\begin{eqnarray*}
m^2_\pm&=&{1\over2}(s+3m_\pi^2-m^2)      \nonumber\\
& &\pm{2\sqrt{s}\over m}q_\pi(m)q_\rho(m)x,   \nonumber\\
q_\rho(m)&=&q(\sqrt{s},m,m_\pi),           \nonumber\\
q_\pi(m)&=&q(m,m_\pi,m_\pi),
\end{eqnarray*}
\cite{ach92}, stands for
the  phase space volume of the $\pi^+\pi^-\pi^0$ final state.
The general form of all propagators including the $\rho(770)$ one,
$D_\rho(m^2)$, whose
imaginary part is determined by the $\pi^+\pi^-$ partial width
$$\Gamma_\rho(m^2)={g^2_{\rho\pi\pi}\over6\pi m^2}q^3_\pi(m),$$
is given in Sec. \ref{susec2c}.

\subsection{Discussing the coupling constants}
\label{susec2b}

Before writing down explicit expressions for the various matrix elements
entering into the matrices above,
let us comment on the coupling constants of the vector mesons with
various final states $f$.

\subsubsection{The final state $\pi^+\pi^-\pi^0$}

The $\rho(770)-\rho^\prime_1-\rho^\prime_2$ sector does not contribute
by G-parity conservation, and 
hence $g_{\rho_if}=0$. The $\omega_i\rho(770)\pi$
coupling constant should be inserted in place of $g_{\omega_if}$.
In particular, $g_{\omega\rho\pi}$  and the OZI suppressed
coupling constant $g_{\varphi\rho\pi}$ are determined from
fitting the data on the present final state and are kept fixed in fitting
the remaining final states. It should be emphasized that the existing data
still cannot distinguish between the two mechanisms of the $\pi^+\pi^-\pi^0$
decay of the $\varphi(1020)$, namely, a sizable $\varphi\omega$ mixing,
$\varphi\to\omega\to\pi^+\pi^-\pi^0$ and the
 direct transition $\varphi\to\pi^+\pi^-\pi^0$
\cite{ach95}. In principle, a careful experimental study of the
$\varphi\omega$ interference minimum in the reaction
$e^+e^-\to\pi^+\pi^-\pi^0$ could discriminate between the above models
\cite{ach93}. However, such subtleties are inessential in the present case,
since both models give a similar behavior of the cross section. So we take
here for  definiteness the purely $s\bar s$ quark content of the
$\varphi(1020)$, thus attributing its $\pi^+\pi^-\pi^0$ decay mode solely
to the direct coupling constant $g_{\varphi\rho\pi}$. The masses and
coupling constants of the $\omega(782)-\varphi(1020)$ complex extracted
from the fit will turn out to be
\begin{eqnarray}
m_\omega&=&783.4^{+2.7}_{-3.0}\mbox{ MeV, }
m_\varphi=1019.8^{+1.6}_{-1.4}\mbox{ MeV, }    \nonumber\\
g_{\omega\rho\pi}&=&14.3\pm1.6\mbox{ GeV}^{-1}\mbox{, }
g_{\varphi\rho\pi}=0.63\pm0.17\mbox{ GeV}^{-1}    \nonumber\\
f_\omega&=&16.6^{+1.7}_{-1.3}.
\label{omphi}
\end{eqnarray}
They coincide, within errors, with the parameters obtained earlier
\cite{ach92}, and so we will not discuss them further.
Note that  we will assume hereafter  the quark model relation
\begin{equation}
f_{\varphi_i}=-{f_{\omega_i}\over\sqrt{2}}
\label{lepcc}
\end{equation}
between the leptonic coupling constants $f_{\omega_i}$ and $f_{\varphi_i}$.

The accuracy of
existing data in the energy range of $\sqrt{s}=1.1-2\mbox{ GeV}$ is
still insufficient (see below) for the introduction of
the nonzero coupling constants  $\varphi^\prime_{1,2}\rho\pi$, hence they
are fixed to zero, so that the  OZI rule breaking in the sectors
which include the heavier excitations, is attributed solely
to the mixing via  the OZI allowed two step processes proceeding through
common decay modes.

\subsubsection{The final state $\omega\pi^+\pi^-$}
The $\rho$-like resonances do not contribute
by G-parity conservation, similar to the  previous case. Our
analysis of pure isovector channels of $e^+e^-$ annihilation \cite{ach97a}
reveals the negligible contribution of the off-mass-shell coupling
$\rho\to\rho\pi^+\pi^-$ in the energy range $\sqrt{s}\geq1 $ GeV.
Guided by the planar quark diagram approach,
a similar $\omega\to\omega\pi^+\pi^-$ coupling is neglected
in the present case.  The coupling constants of the $\varphi$-like
resonances to the state under consideration
are suppressed by the OZI rule and hence can be
set to zero, bearing in mind the poor accuracy of the present data.
Note also that the $\omega\pi^+\pi^-$ mode takes into account
effectively the $b_1\pi$, etc., modes. In fact, the chain
$\omega^\prime_{1,2}\to b_1(1235)\pi\to\omega\pi^+\pi^-$ includes the decay
of the axial $b_1$ whose decay amplitude contains two independent
partial waves. This results, in general, in a structureless angular
distribution of final pions and could be modelled by the effective
pointlike $\omega\pi^+\pi^-$ vertex, which includes also possible
intermediate states containing the scalarlike mesons, $\omega^\prime\to
\omega\sigma\to\omega\pi^+\pi^-$.

The $\varphi_{1,2,3}\to\varphi\pi\pi$ coupling constant is expected to be
suppressed due to the OZI rule and hence is omitted. This guess is supported
by the fact that the final state $\varphi\pi\pi$ is not observed in $e^+e^-$
annihilation \cite{pdg}.

\subsubsection{The final state $K^+K^-$}

The contribution of the $\rho$- and $\omega$-like resonances is
taken into account via  SU(3) relations for their coupling, assuming
a $q\bar q$ quark content:
\begin{eqnarray}
g_{\rho^0_{1,2,3}K^+K^-}&=&-{1\over\sqrt{2}}g_{\varphi_{1,2,3}K^+K^-},
\nonumber\\
g_{\omega_{1,2,3}K^+K^-}&=&-{1\over\sqrt{2}}g_{\varphi_{1,2,3}K^+K^-},
\label{sukk}
\end{eqnarray}
and the SU(2) related to the above. As a further fit shows,
\begin{equation}
g_{\varphi K^+K^-}=4.7\pm0.5,
\label{phikk}
\end{equation}
and so we will not discuss this coupling constant further anymore.
The parameters of the $\rho$ excitations
are chosen as follows. The analysis \cite{ach97a} of these excitations
gives a number of variants of the best description of the specific final
state, and the parameters extracted from various final states
agree within errors.
We plot the $\rho(770)+\omega(782)+\varphi(1020)+\rho^\prime_1+\rho^\prime_2$
resonance contribution  to the production
cross section of the $K^+K^-$ final state and convince ourselves that,
surprisingly, the $\rho^\prime_{1,2}$ parameters from
the variant of the description of the
reaction $e^+e^-\to\pi^+\pi^-\pi^0\pi^0$ (with the subtraction of the
$\omega\pi^0$ events) \cite{ach97a} falls closer to the data
\cite{olya,dm2_kk}
than other variants do, and by that reason this variant is adopted in the
present case, hereafter dubbed the set $A$.
In the meantime, another  set of the $\rho^\prime_{1,2}$ parameters from
\cite{ach97a} is briefly discussed at the proper place below.

\subsubsection{The final state $K^0_SK^\pm\pi^\mp$}
This final state originates from the $K^{\ast0}\bar K^0+\bar K^{\ast0}K^0$
intermediate state, so that the production cross sections  are
related as
$\sigma(K^0_SK^+\pi^-)={1\over2}\sigma(K^{\ast0}\bar K^0)$, analogously
for the charge conjugated states . Hence,
the coupling constant $g_{VK^\ast\bar K}$ should be inserted instead
of $g_{Vf}$, and a factor of 1/2 appears in the
corresponding expression for $P_f$.
The coupling constants of the $\rho$-like,  $\omega$-like,
and  $\varphi$-like resonances   are supposed to obey the $q\bar q$ model
relations
\begin{eqnarray}
g_{\rho_{1,2,3}K^{\ast+}\bar K^-}&=&{1\over2}g_{\omega_{1,2,3}\rho\pi},
\nonumber\\
g_{\omega_{1,2,3}K^{\ast+}\bar K^-}&=&{1\over2}g_{\omega_{1,2,3}\rho\pi},
\nonumber\\
g_{\varphi_{1,2,3}K^{\ast+}\bar K^-}&=&{1\over\sqrt{2}}
g_{\omega_{1,2,3}\rho\pi},
\label{relations}
\end{eqnarray}
and the SU(2) related to them.  The parameters
of the $\rho$-like excitations are the same as for the final state
$K^+K^-$.

\subsubsection{Final state $K^{\ast0}K^-\pi^+
(\bar K^{\ast0}K^+\pi^-)$}

The production amplitude of this final state includes, in principle,
both the effective pointlike, in the sense explained earlier in the case
of the $\omega\pi^+\pi^-$ decay channel,  $V\to K^{\ast0}K^-\pi^+$, and
the triple
vector, $V\to K^{\ast0}\bar K^{\ast0}$, vertices. The latter is the SU(3)
related to the vertex $\rho_i\to\rho^+\rho^-$.  It was shown earlier
\cite{ach97a}, that the contribution of the $\rho^0\to\rho^+\rho^-$ tail
is negligible at $\sqrt{s}>1$ GeV, while the $\rho^\prime_{1,2}\to\rho^+
\rho^-$ coupling constants are compatible with zero. Guided by SU(3), it is
reasonable to neglect the triple vector couplings in the present case, too.
Note that a zero isospin of the vector mesons involved in the
effective pointlike
four-particle vertices allows us
to relate various charge combinations of pions
and kaons, resulting in  ratio of the coupling constants:
\begin{eqnarray}
K^{\ast0}K^-\pi^+:\bar K^{\ast0}K^+\pi^-:K^{\ast-}K^+\pi^0:
K^{\ast+}K^-\pi^0  \nonumber\\
=1:1:{1\over\sqrt{2}}:{1\over\sqrt{2}}.
\label{rel1}
\end{eqnarray}
Hence, we retain here only the coupling constants
$g_{\omega^\prime_{1,2}K^{\ast0}K^-\pi^+}$,
$g_{\varphi^\prime_{1,2}K^{\ast0}K^-\pi^+}$  and keep them free.
The $\rho^\prime_{1,2}\to K^{\ast}\bar K\pi$
couplings contain two isotopic states of the $K^\ast\bar K+\bar K^\ast K$
system, $I=0,1$. Then the ratios of the coupling constants of the neutral
$\rho^\prime_{1,2}$ states to various $K^\ast\bar K\pi$
charge states are expressed through the coupling constants
$c^{(I)}_{\rho^\prime_{1,2}}$ with 
definite isospin according to
\begin{eqnarray}
K^{\ast+}K^-\pi^0:K^{\ast0}\bar K^0\pi^0:K^{\ast+}\bar K^0\pi^-
:K^{\ast0}K^-\pi^+       \nonumber\\
=c^{(0)}_{\rho^\prime_{1,2}}+{c^{(1)}_{\rho^\prime_{1,2}}\over\sqrt{2}}
:c^{(0)}_{\rho^\prime_{1,2}}-{c^{(1)}_{\rho^\prime_{1,2}}\over\sqrt{2}}:
c^{(1)}_{\rho^\prime_{1,2}}:c^{(1)}_{\rho^\prime_{1,2}},
\label{rhokstk}
\end{eqnarray}
and the charge conjugated to the above. These couplings
were neglected in \cite{ach97a}. In principle, they should be
included in the future, after obtaining  good consistent data on various
channels.
In the following analysis of the final states containing 
strange mesons, we will set  upper bounds on the
$\rho^\prime_{1,2}\to K^\ast\bar K\pi$ couplings with  definite isospin
of the $K^\ast\bar K$ states.

\subsection{Propagators and the nondiagonal polarization operators}
\label{susec2c}

The propagator of the bare vector meson $V$ and the imaginary part of the
nondiagonal polarization
operator describing the mixing between the bare states $V_i$ and $V_i$
($V=\omega, \varphi$) are, respectively,
\begin{equation}
D_{V_i}\equiv D_{V_i}(s)=m^2_{V_i}-s-i\sqrt{s}\Gamma_{V_i}(s)
\label{prop}
\end{equation}
and
\begin{eqnarray}
\mbox{Im}\Pi_{V_iV_j}(s)&=&\sqrt{s}\left(g_{V_i\rho\pi}g_{V_j\rho\pi}
P_{\pi^+\pi^-\pi^0}\right.        \nonumber\\
& &+2g_{V_iK^+K^-}g_{V_jK^+K^-}P_{K^+K^-}   \nonumber\\
& &\left.+4g_{V_iK^{\ast+} K}g_{V_jK^{\ast+}K^-}P_{K^0_SK^+\pi^-}
\right.        \nonumber\\
& &\left.+6g_{V_iK^{\ast0}K^-\pi^+}g_{V_jK^{\ast0}K^-\pi^+}
P_{K^{\ast0}K^-\pi^+}\right.       \nonumber\\
& &\left.+g_{V_iV_1\pi^+\pi^-}g_{V_jV_1\pi^+\pi^-}W_{VP\pi}\right),
\label{imag}
\end{eqnarray}
where $W_{VP\pi}\equiv W_{VP\pi}(\sqrt{s},m_{V_1},m_\pi)$. Here
the factors of 2, 4, and 6 multiplying, respectively,
the $K\bar K$,  $K^\ast\bar K$, and $K^\ast\bar K\pi$ contributions
allow for various charge combinations,
taken with the proper SU(2) coefficients [see Eq. (\ref{rel1})], and the
phase space factors of different final states are given in Eq.
(\ref{eq2.5}). The width
of the bare state $V_i$ can be represented, in this notation, as
\begin{equation}
\Gamma_{V_i}(s)=\mbox{Im}\Pi_{V_iV_i}(s)/\sqrt{s}.
\label{width}
\end{equation}
The sector $V=\rho$ was described earlier \cite{ach97a}. Here we should
add the partial width
\begin{eqnarray}
\Gamma(\rho^\prime_{1,2}\to K^\ast\bar K\pi+\mbox{c.c})&=&
(4c^{(0)2}_{\rho^\prime_{1,2}}+6c^{(1)2}_{\rho^\prime_{1,2}})
       \nonumber\\
& &\times W_{VP\pi}(\sqrt{s},m_{K^\ast},m_K)
\label{parwid}
\end{eqnarray}
to the full width of the $\rho^\prime_{1,2}$ and the contribution
$$\sqrt{s}\left(4c^{(0)}_{\rho^\prime_1}c^{(0)}_{\rho^\prime_2}+
6c^{(1)}_{\rho^\prime_1}c^{(1)}_{\rho^\prime_2}\right)
W_{VP\pi}(\sqrt{s},m_{K^\ast},m_K)$$
to the imaginary part of the nondiagonal polarization operator
$\Pi_{\rho^\prime_1\rho^\prime_2}$ \cite{ach97a}.

The expressions for the partial widths could include the energy-dependent
factors $C_f(s)$ which,
analogously to the well known Blatt-Weiskopf centrifugal factors, are
aimed at  restricting a too fast growth of the partial widths with the energy
rise. They are somewhat arbitrary under the demand of $\sqrt{s}\Gamma(s)\to$
const at $\sqrt{s}\to\infty$. In practice, the only mode with a strong
dependence is
the vector (V)+pseudoscalar (P) one,  and our choice for the factor
multiplying corresponding coupling constant is
\begin{equation}
C_{VP}(s)={1+(R_{VP}m_0)^2\over1+(R_{VP}\sqrt{s})^2},
\label{blatt}
\end{equation}
where $m_0$ is the mass of the resonance and $R_{VP}$ is the so-called range
parameter.

Contrary to the imaginary parts fixed by the unitarity relation, the real
parts of {\it all} nondiagonal polarization operators cannot
be evaluated at present and hence should be taken as free
parameters. However, some information about the mass spectrum of the ground
state mesons can provide a reasonable guess about the real parts
of the nondiagonal polarization operators describing the mixing of the ground
state mesons with the heavier ones. In fact, it was shown earlier
\cite{ach92} in the case of two mixed states 1 and 2 that the masses of both
these states acquire shifts in the opposite directions. In particular,
the shift of the lower state is
\begin{equation}
\delta m_1\simeq-\mbox{Re}{\Pi^2_{12}(m^2_1)/2m_1
\over m^2_2-m^2_1-im_1[\Gamma_2(m^2_2)-\Gamma_1(m^2_1)]},
\label{mshift}
\end{equation}
and it can be large. However, a fit by the minimization of the $\chi^2$
function fixes only the combination $m_1+\delta m_1$. Hence, it is natural
to assume that the dominant contribution to the mass renormalization,
Eq. (\ref{mshift}), coming from $(\mbox{Re}\Pi_{12}^2)$ is already subtracted,
so that the mass of the lower state minimizing $\chi^2$ differs from the
actual position of  peak 1 in the cross section by a quantity quadratic
in Im$\Pi_{1,2}$. Then the minimum of $\chi^2$ is provided by the values
of Re$\Pi_{1,2}$ falling close to zero, of course, with very large error
bars. So it is reasonable to fix the latter to zero from the very start.
These considerations, justifiable in the case of the ground state mesons
$\rho(770)$, $\omega(782)$, and $\varphi(1020)$ whose $q\bar q$ nature is
firmly established, cannot be applied to the higher excitations. The latter
may
contain an appreciable admixture of an exotic component like $q\bar qg$,
$q^2\bar q^2$, etc. \cite{kalash}, and 
so it is a matter of principle to extract
from the data the parameters of unmixed states. By this reason the real
parts of the nondiagonal polarization operators describing the mixing among
heavier excitations should be kept free. We consider them to be independent
of energy.

\section{Results}
\label{sec3}

The procedure of the extraction of the resonance parameters is the same
as in \cite{ach97a}. We fit the data on each reaction Eqs. (\ref{dm2_3pi})$-$
(\ref{dm2_kstkpi}) separately, by minimizing the $\chi^2$ function.
In principle, the specific set of  parameters
giving the best description of the specific cross section unnecessarily
gives a good description of other channels. Hence, final choice is
made on the demand that the sets of the parameters obtained from various
channels should differ by no more than one to two standard deviations. This
looks reasonable, especially if one bears in mind the desirable possibility
of gathering good consistent data on different channels and on the single
facility.

It should be emphasized that the usual representation of the resonance
parameters as  masses and  partial widths evaluated at these masses
is inadequate in the case of strongly mixed resonances and the strong
energy dependence of the partial widths. As will be clear later on, actual
peaks in the cross sections are displaced considerably from the input masses
of the heavier excitations. This is the reason for our choice
of the masses and {\it coupling constants} of bare states, not their partial
widths, to represent the
results. Furthermore, a large number of free parameters pushes us to invoke
some hypotheses on the relations between the coupling constants.
The assumption adopted in the present paper is the $q\bar q$ nature of the
isoscalar excitations. Corresponding relations among the hadronic coupling
constants are given in Sec. \ref{susec2b}.

Our results  are collected in Tables \ref{tab1}, \ref{tab2},
and in Fig. \ref{fig1}$-$ \ref{fig5}. In Table \ref{tab3} we quote the values
of $\chi^2/n_{\rm DOF}$ for each channel considered in the work.
It is seen that  all the parameters agree within the large error bars.
The range parameter $R_{\rho\pi}$ for the $\rho\pi$ decay mode
[and the SU(3) related
to it] turns out to be $0.4\pm1.0\mbox{ GeV}^{-1}$ in the case of the
$\omega(782)$, and is not fixed by the fit in the case of the
$\omega^\prime_1$. We choose it to be
zero for the latter excitation and for the
$\omega^\prime_2$ one, too.

Let us comment on the
visible disagreements in Tables \ref{tab1} and \ref{tab2}.
First, the small, compared to the other, value of
$\Gamma_{\omega^\prime_{1,2}e^+e^-}$ extracted from the $K^+K^-$
data is an artifact of our particular choice of  $\rho^\prime_{1,2}$
resonance parameters. Another choice \cite{ach97a},
with the parameters extracted from
the $e^+e^-\to\pi^+\pi^-\pi^+\pi^-$ data hereafter dubbed as the set $B$,
gives better values,
$\Gamma_{\omega^\prime_1e^+e^-}=(20^{+60}_{-20})\times10^{-2}$ keV and
$\Gamma_{\omega^\prime_2e^+e^-}=(11^{+19}_{-11})\times10^{-2}$ keV. This
emphasizes the necessity of obtaining consistent data about various final
states in $e^+e^-$ annihilation.
The visible disagreement  of the central value of the leptonic width
$\Gamma_{\omega^\prime_2e^+e^-}$
extracted from the reaction  (\ref{dm2_kstkpi}) is due to
the following. First, the error bars are so large that the disagreement is
statistically insignificant. Second, the threshold of the reaction,
Eq. (\ref{dm2_kstkpi}), 1.53 GeV, is so high that the inclusion of additional
multiparticle decay modes may be necessary, which could  change the
result towards better values. We postpone this task until more satisfactory
experimental data will appear.

Notice that the peak positions of the heavier excitations are displaced
towards
lower values from the bare masses of resonances. The same phenomenon was
observed in the case of isovector excitations \cite{ach97a} and is
due, predominantly, to the growth of the partial widths with the energy,
\begin{equation}
\delta   m_{V^{\prime}_{1,2}}\sim -\Gamma(s){d\Gamma\over d\sqrt{s}}
(\sqrt{s}=m_{V^\prime_{1,2}}).
\label{shift}
\end{equation}

Unfortunately, available data do not put any serious restrictions on the real
parts of the nondiagonal polarization operators. The minimization procedure
points to possible nonzero values which are quoted in Table \ref{tab2}, but
the error bars are very large.

The best upper bounds on the  $\rho^\prime_{1,2}K^\ast\bar K\pi$
couplings $c^{(I)}_{\rho^\prime_{1,2}}$ with
definite isospin of the $K^\ast\bar K$ system are as follows:
i)  $|c^{(0)}_{\rho^\prime_1}|<1500$ and $|c^{(0)}_{\rho^\prime_2}|<1100$
come from fitting the
cross section of the reaction $e^+e^-\to K_SK^\pm\pi^\mp$;
ii)  $|c^{(1)}_{\rho^\prime_2}|<390$ and $|c^{(1)}_{\rho^\prime_2}|<210$
come from fitting the
cross section of the reaction $e^+e^-\to K^{\ast0}K^-\pi^+$.

\section{Discussion}
\label{sec4}

The nonrelativistic quark model (NRQM) for  bound states of light
quarks cannot be easily justified in view of the expectedly  large QCD and
relativistic corrections. Yet the remarkable agreement of NRQM predictions
with the data in the physics of light quarks is impressive, and by this
reason the NRQM provides a conventional reference frame  for representing
the results of numerous analyses. In particular,  conclusions about
the possible non-$q\bar q$ component \cite{kalash,clegg94} are based on
a comparison of calculations  \cite{page95}
with the ratios of coupling constants based on the NRQM. So we also
will follow here the custom of expressing the results in terms of
the interquark potential, bound state wave functions, etc., bearing in mind
that the extent of the reliability to the models of such a kind is supported
by their effectiveness in known cases rather than by a firm theoretical
basis.

As is known, the leptonic widths, Eq. (\ref{eq2.4}), are sensitive to
the behavior of the wave function of the bound $q\bar q$ state at the
origin \cite{weiskopf,novikov}:
\begin{eqnarray}
\Gamma(^3S_1\to e^+e^-)&=&{4\alpha^2Q^2_V\over m^2}|R_S(0)|^2,  \nonumber\\
\Gamma(^3D_1\to e^+e^-)&=&{200\alpha^2Q^2_V\over m^6}
|R^{\prime\prime}_D(0)|^2,
\label{l_width}
\end{eqnarray}
where $R_L(0)$ is the radial wave function at the origin of the $q\bar q$
bound state with  angular momentum $L$. $Q_V$ is related to the quark
content of the vector meson
$V=\rho_i\mbox{, }\omega_i\mbox{, }\varphi_i$ and reads
$1/\sqrt{2}\mbox{, }1/3\sqrt{2}\mbox{, }-1/3$ for, respectively,
$\rho_i=(u\bar u-d\bar d)/\sqrt{2}\mbox{, }
\omega_i=(u\bar u+d\bar d)/\sqrt{2}\mbox{, }\varphi_i=s\bar s$, assuming a
$q\bar q$ nature of the heavier excitations. To make the comparison easier,
we quote the magnitudes of the wave function and the second derivative at
the origin averaged over the channels under consideration. The results for
$\rho$-like excitations are evaluated with the help of \cite{ach97a}.
One obtains, for the $\rho$-like excitations,
\begin{eqnarray}
|R_S(0,m_{\rho^\prime_1})|^2&=&110^{+30}_{-20}\times10^{-3}\mbox{GeV}^3,
\nonumber\\
|R_S(0,m_{\rho^\prime_2})|^2&=&(140\pm20)\times10^{-3}\mbox{GeV}^3,
\nonumber\\
|R^{\prime\prime}_D(0,m_{\rho^\prime_1})|^2&=&
80^{+20}_{-10}\times10^{-4}\mbox{GeV}^7,
\nonumber\\
|R^{\prime\prime}_D(0,m_{\rho^\prime_2})|^2&=&
(300\pm50)\times10^{-4}\mbox{GeV}^7,
\label{wfun_rho}
\end{eqnarray}
to be compared to
$|R_S(0,m_\rho)|^2=37\times10^{-3}\mbox{GeV}^3$.
As is pointed out in Sec. \ref{sec3}, the amplitude of the reaction
$e^+e^-\to K^{\ast0}K^-\pi^+$ seems to be affected by the multiparticle
intermediate states neglected in the present analysis, and so we exclude it
from  averaging in the case of the isoscalars. One gets
\begin{eqnarray}
|R_S(0,m_{\omega^\prime_1})|^2&=&40^{+620}_{-40}(140^{+620}_{-140})
\times10^{-4}\mbox{GeV}^3,
\nonumber\\
|R_S(0,m_{\omega^\prime_2})|^2&=&430^{+1420}_{-250}(400^{+1080}_{-200})
\times10^{-4}\mbox{GeV}^3,
\nonumber\\
|R^{\prime\prime}_D(0,m_{\omega^\prime_1})|^2&=&
30^{+960}_{-30}(110^{+980}_{-110})\times10^{-5}\mbox{GeV}^7,
\nonumber\\
|R^{\prime\prime}_D(0,m_{\omega^\prime_2})|^2&=&
140^{+570}_{-100}(120^{+430}_{-80})\times10^{-4}\mbox{GeV}^7,
\label{wfun_om}
\end{eqnarray}
to be compared to
$|R_S(0,m_\omega)|^2=33\times10^{-3}\mbox{GeV}^3$.
Here the numbers in the parentheses refer to the set $B$ of the parameters
of the $\rho$-like excitations mentioned earlier. Note that within error
bars the numerical characteristics of the $q\bar q$ structure of the
$\rho$-like and $\omega$-like excitations
are coincident, thus supporting their assignment to the same nonet.

A realistic interquark potential could include the sum of a Coulomb-like
one, with a running QCD coupling constant and a confining potential
\cite{godfrey85}. However, because the errors in extracting the wave functions
and second derivatives at the origin are still too large, one cannot draw
any definite conclusions about the parameters of the potential or to verify
the usual assignments $\rho^\prime_1\equiv\rho(1450)\sim 2^3S_1$ and
$\rho^\prime_2\equiv\rho(1600)\sim 1^3D_1$, similar for
$\omega^\prime_1$ and $\omega^\prime_2$.

The present study differs basically from that of Ref.
\cite{clegg94}, where the simplest Breit-Wigner amplitude, with the neglect of
mixing among the states, was used for fitting the cross sections.
So our conclusions are different from those of
\cite{kalash,clegg94,page95}
at the point that the $q\bar q$ nature of
heavier resonances is not excluded by existing data. The $q\bar q$ model
relation, Eq. (\ref{lepcc}), 
is fixed here from the very start, while the ratio
$\Gamma(\omega^\prime_i\to e^+e^-)/\Gamma(\rho^\prime_i\to e^+e^-)$
spreads from zero to 0.3 in all the fits and does not contradict to the
$q\bar q$ ratio $\sim1/9$. The vector-pseudoscalar and pseudoscalar-
pseudoscalar couplings are also related via the $q\bar q$ model in our work.
It should be recalled that our previous analysis \cite{ach97a} revealed the
ratio $B(\rho^\prime_1\to2\pi^+2\pi^-)/B(\rho^\prime_1\to\omega\pi^0)$
to be consistent with zero. This is in a  contrast with the hypothesis of the
hybrid admixture \cite{kalash,clegg94,page95} which predicts the dominance
of final states containing the $L=1$ mesons \cite{page95}.
Pseudoscalar-pseudoscalar final states appear to
be suppressed in our analysis,
in qualitative agreement with the considerations of Ref. \cite{page95}.
Note, however, that there are reasons to expect the suppression of these
final states for pure $q\bar q$ mesons \cite{odor}.

\section{Conclusion}
\label{sec5}

The data on the isoscalar heavier excitations existing now are still
of poor accuracy, not only in that the errors of the extracted parameters of
these excitations are large, but in that these data are insufficient to
discriminate between possible dynamical models of the resonances with
masses greater than 1 GeV. In fact, we find an alternative variant
of the description which does not at all demand the existence of 
$\omega^\prime_1$ and $\varphi^\prime_1$ resonances, however, at the
expense of abandoning the $q\bar q$ model relations among  strong
coupling constants. The corresponding curves look even better than those
shown in Figs. \ref{fig1}$-$\ref{fig5} in that  $\omega^\prime_1$
and $\varphi^\prime_1$ peaks, which seem accidental, are absent in this
variant.

Looking at the curves in the present paper convinces us that
fitting the scarce data with  expressions containing many free parameters
can bring one to the trap of low $\chi^2$ when the curve goes through the
central points, each possessing large error bars. A narrow structure at
$\sqrt{s}\simeq 1.5$ GeV seen in Figs. \ref{fig1} $-$ \ref{fig3} and
\ref{fig6}, is due to the
$\varphi^\prime_1$ resonance. Nonzero coupling constants of the
$\varphi^\prime_1$ resulting in its narrow width possess very large errors
which make the former to be consistent with zero.
However, simply dropping them
makes $\chi^2$ considerably larger. Nevertheless, we include just
the present variant because it is coherent with the variant of the
description of the isovector channels \cite{ach97a} based on the picture
of two heavier resonances $\rho^\prime_1$ and $\rho^\prime_2$ \cite{donnach87}.
An additional illustration of such a trap is the channel $e^+e^-\to K_LK_S$.
The curve with  parameters obtained from fitting the reaction
$e^+e^-\to K^+K^-$ and with the proper reversing of sign of the isovector
contribution goes through four low energy OLYA points \cite{ivanov82} but
fails to describe the higher energy DM1 data \cite{mane81} consisting of eight
experimental points. When we fit the data \cite{ivanov82,mane81} on neutral
kaons, we obtain $\chi^2$ as low as 5. The curve shown in Fig. \ref{fig6}
goes through almost each
experimental point, yet looks unnatural
in view of very large error bars.
The only way of escaping a trap of this sort and of proving or disproving
any particular model of  resonances in the mass range $1<m<2.5$ GeV is to
collect good consistent data on all relevant channels. The ranges of
admissible resonance parameters found in the present paper and in the
earlier one \cite{ach97a} will be hopefully useful.

\acknowledgments

We would like to thank G.~N.~Shestakov for discussions.
The present work was supported in part by  grant INTAS-94-3986.

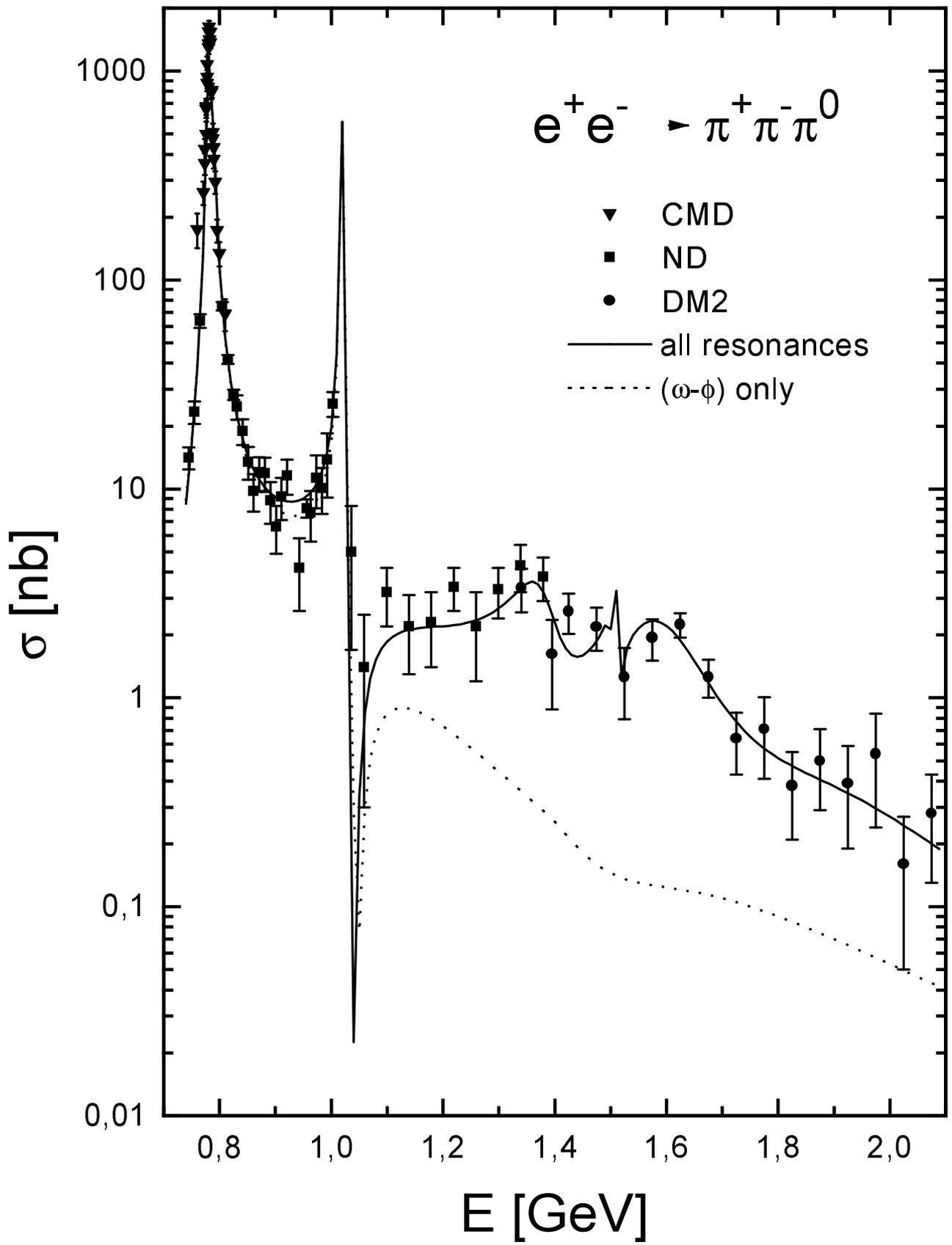
\begin{figure}
\caption{Cross section of the reaction $e^+e^-\to\pi^+\pi^-\pi^0$.
The data are from ND \protect\cite{nd},
DM2 \protect\cite{dm2_3pi}, CMD \protect\cite{cmd}. \label{fig1}}
\end{figure}
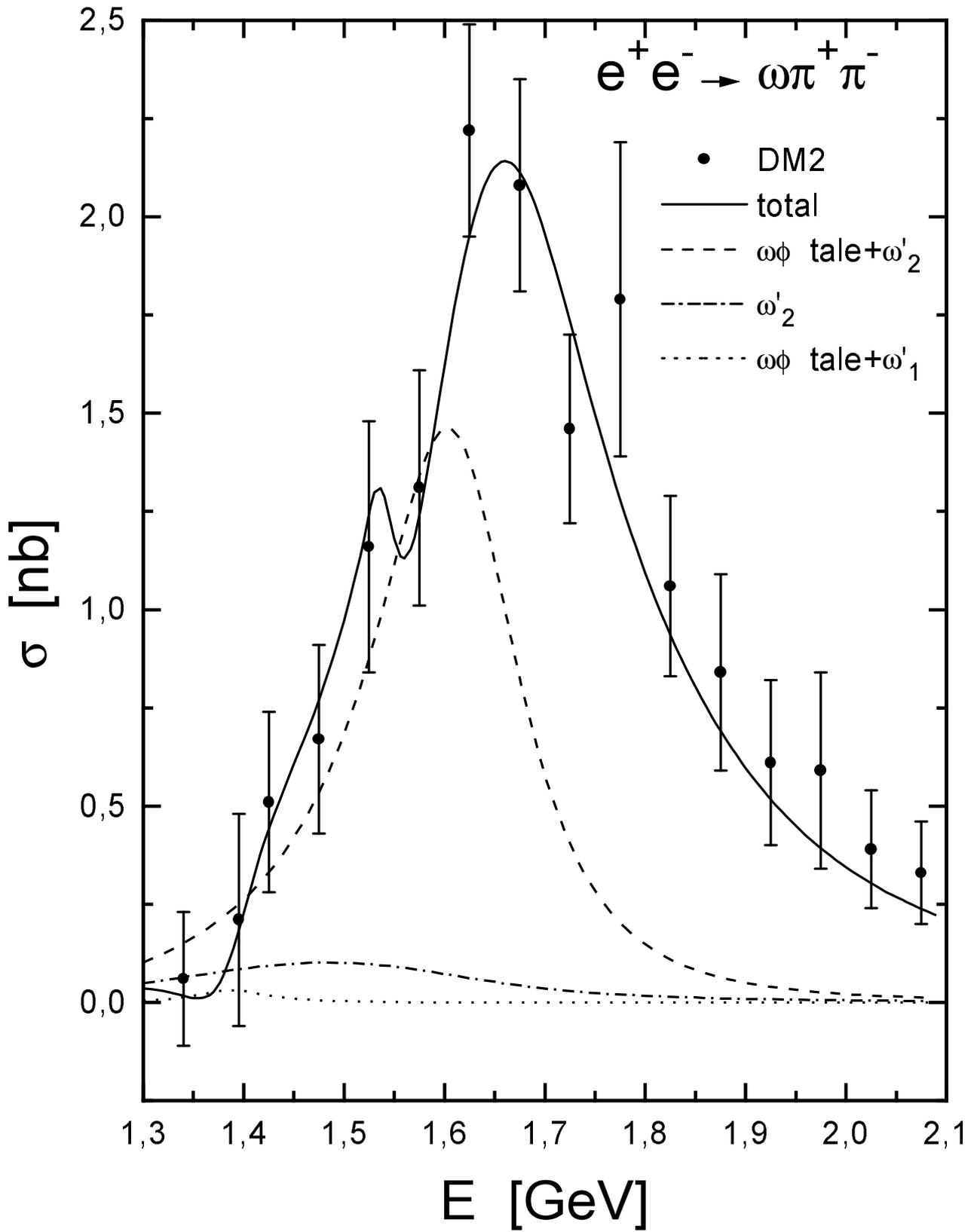
\begin{figure}
\caption{Cross section of the reaction $e^+e^-\to\omega\pi^+\pi^-$.
The data are from DM2 \protect\cite{dm2_3pi}.
\label{fig2}}
\end{figure}
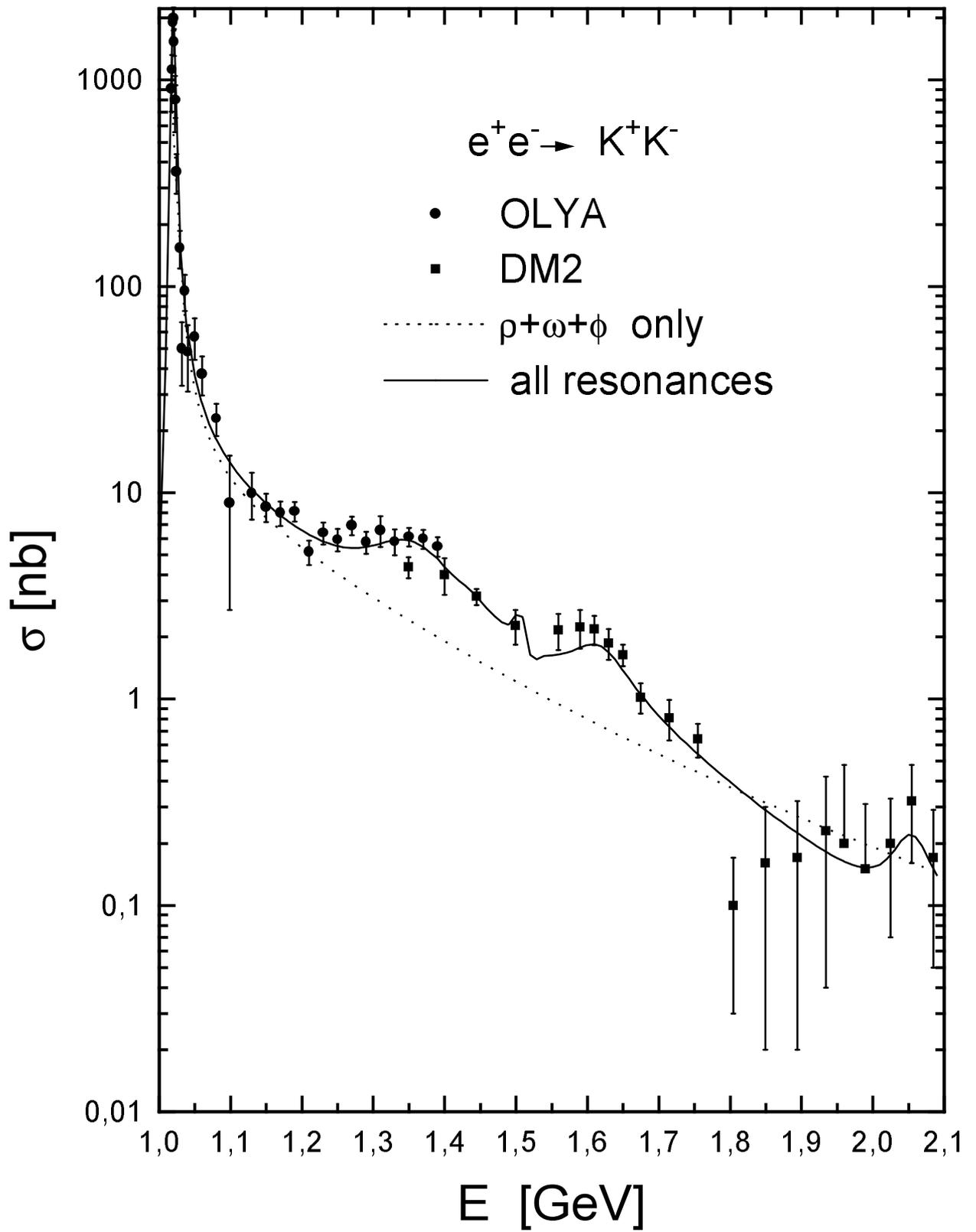
\begin{figure}
\caption{Cross section of the reaction $e^+e^-\to K^+K^-$.
The data are from OLYA \protect\cite{olya}, DM2 \protect\cite{dm2_kk}.
Lower error bars of two experimental points around 1.95 GeV are not
shown because their lowest values are below 0.01 nb.
\label{fig3}}
\end{figure}
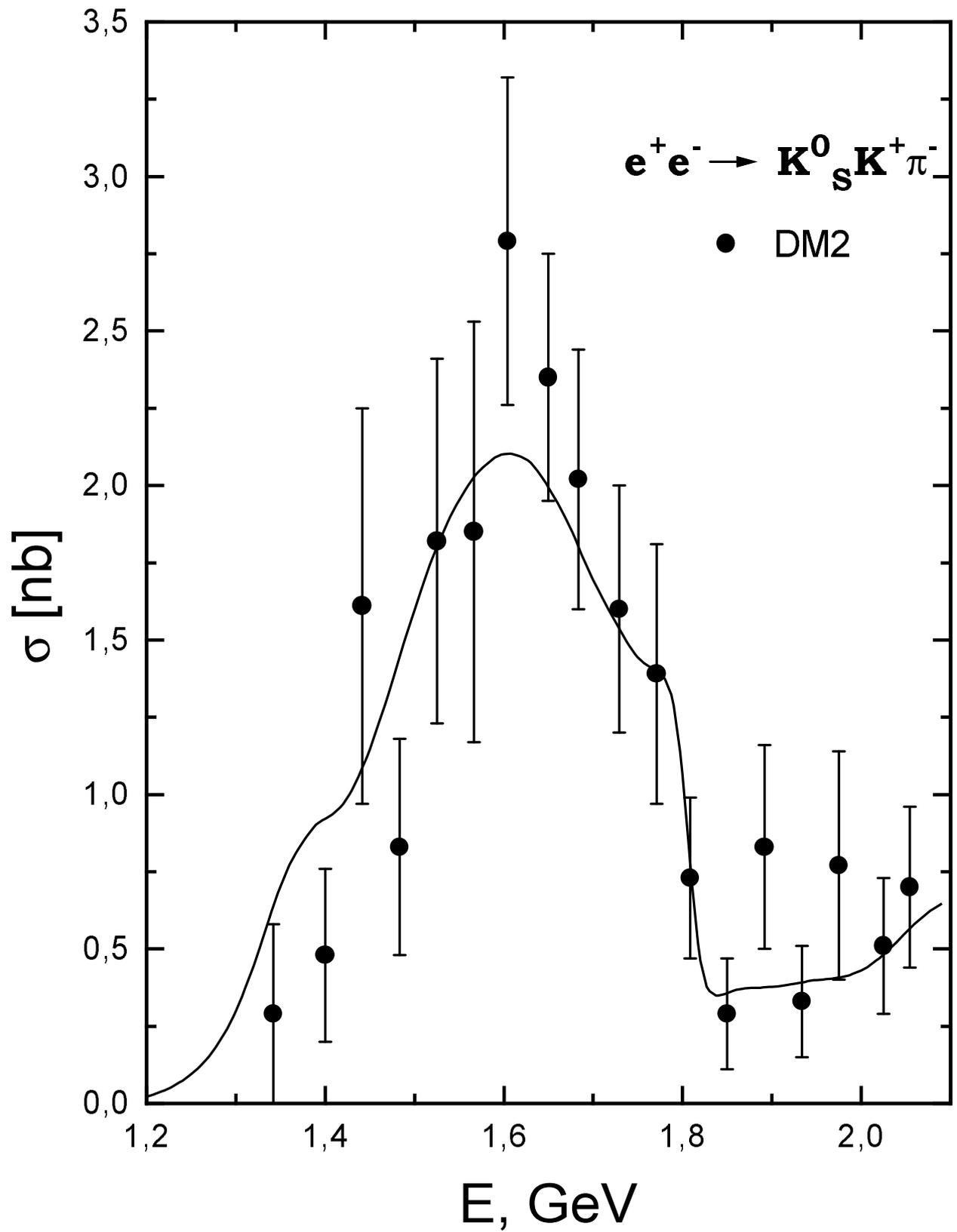
\begin{figure}
\caption{Cross section of the reaction $e^+e^-\to K^0_SK^\pm\pi^\mp$.
The data are from DM2 \protect\cite{dm2_kstk}.
\label{fig4}}
\end{figure}
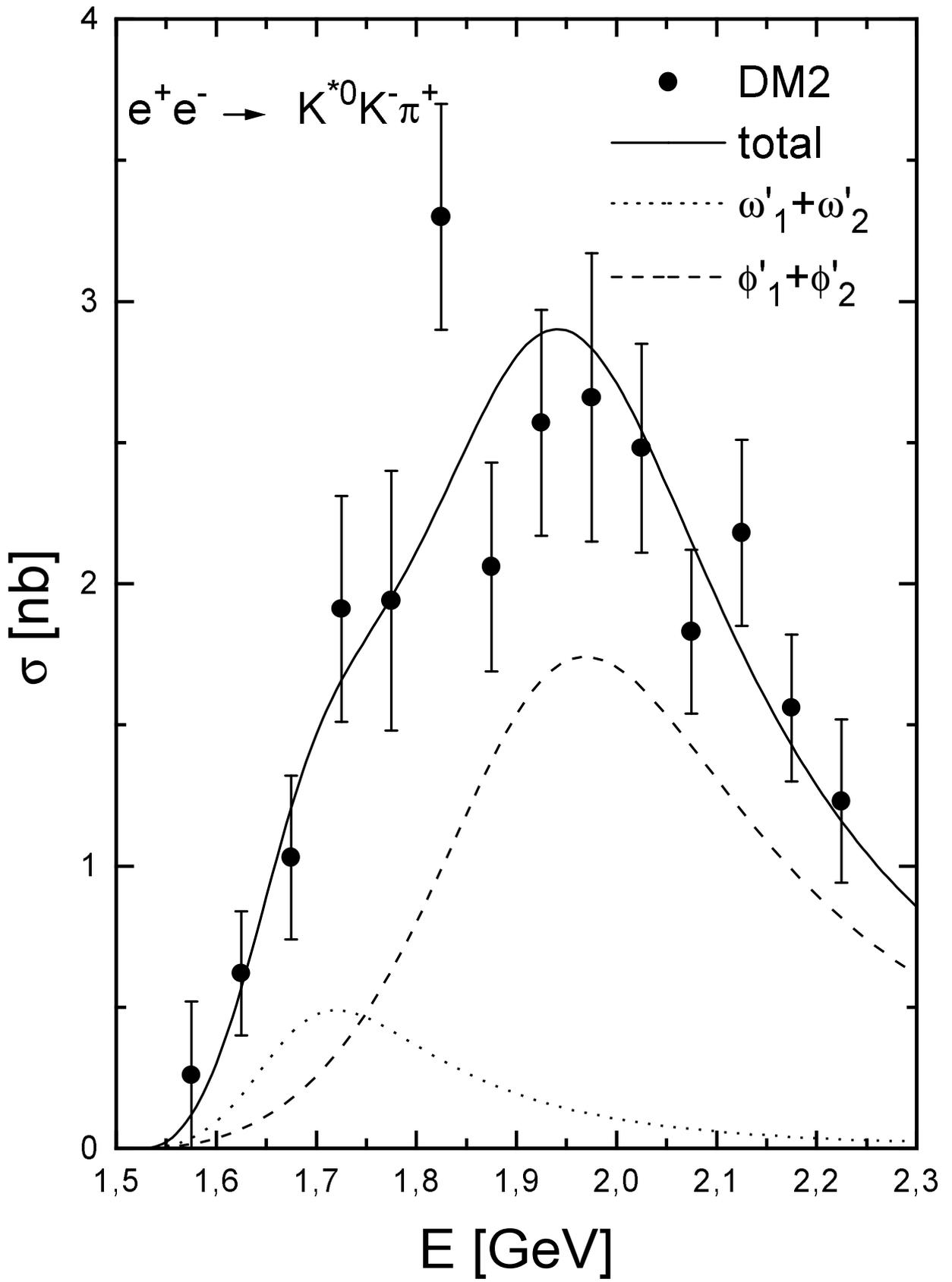
\begin{figure}
\caption{Cross section of the reaction $e^+e^-\to K^{\ast0}K^-\pi^+$.
The data are from DM2 \protect\cite{dm2_kstkpi}.
\label{fig5}}
\end{figure}
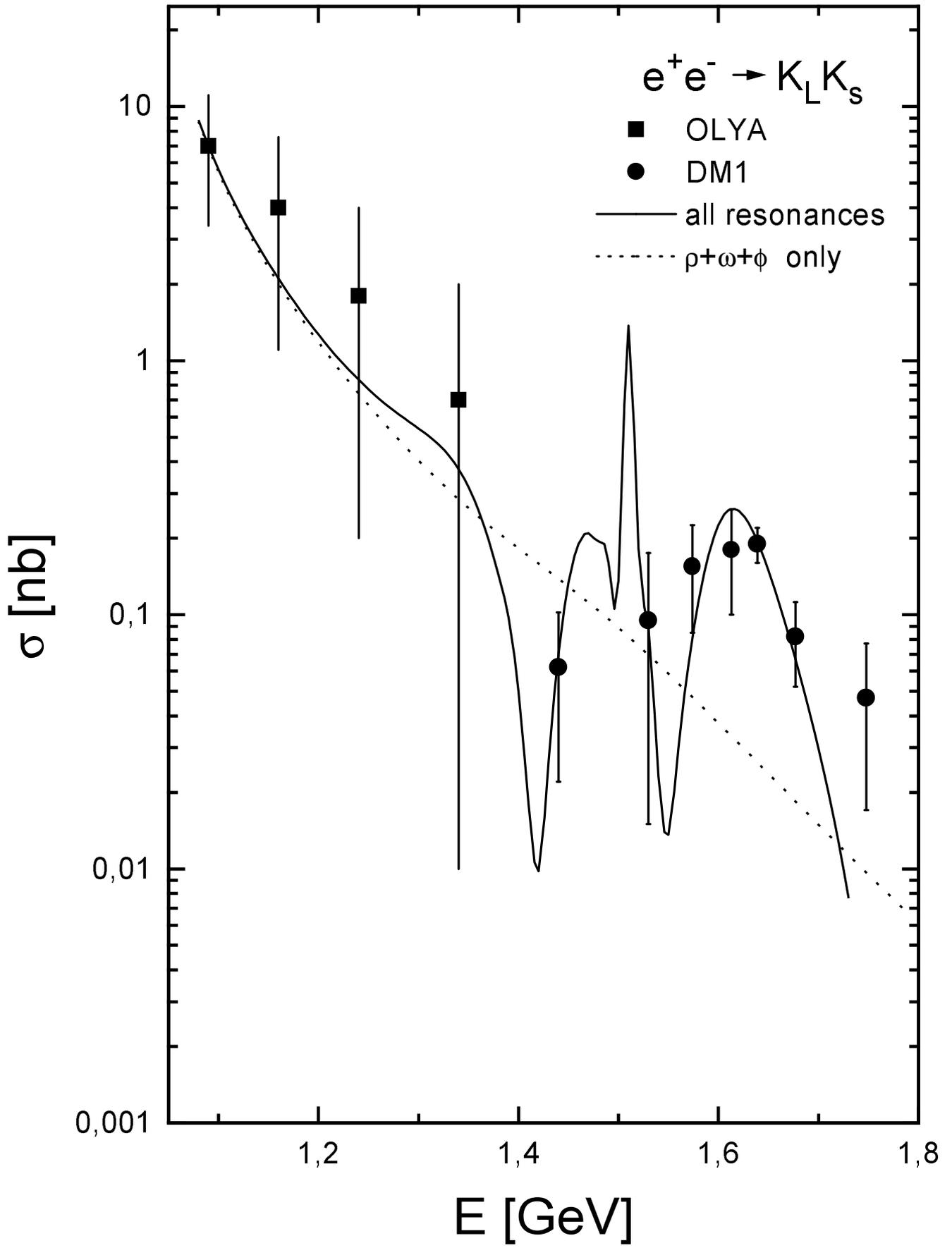
\begin{figure}
\caption{Cross section of the reaction $e^+e^-\to K_LK_S$. Since the data
of OLYA \protect\cite{ivanov82} and DM1 \protect\cite{mane81} have large
error bars, the  parameters of the resonances extracted are also have large
errors and hence are omitted from 
Tables \protect\ref{tab1} and \protect\ref{tab2}.}
\label{fig6}
\end{figure}
\newpage
\widetext
\begin{table}
\caption{The parameters of the $\omega^\prime_1$ and $\varphi^\prime_1$
resonances giving the best description of the data on various final states.
The error bars are
determined from the $\chi^2$ function. The symbol $\sim$ means that $\chi^2$
varies insignificantly upon large variations around the
corresponding parameter.
The modulus of the 
corresponding  parameter is implied in the case of the upper
bound.\label{tab1}}
\begin{tabular}{cccccc}
final state&$\pi^+\pi^-\pi^0$&$\omega\pi^+\pi^-$&$K^+K^-$&$K^0_SK^\pm\pi^\mp$
&$K^{\ast0}K^-\pi^+$\\
\tableline
$m_{\omega^\prime_1}[\mbox{GeV}]$&$1.4^{+0.1}_{-0.2}$&$\sim1.4$&$\sim1.46$&
$1.5^{+0.3}$\tablenote{$\chi^2$ is insensitive to lower values of the mass}
&$1.4^{+0.7}$\tablenotemark[1]\\
$\Gamma_{\omega^\prime_1e^+e^-}[\mbox{keV}]$&$5^{+18}_{-5}\times10^{-2}$
&$<1\times10^{-2}$&$6^{+200}_{-6}\times10^{-3}$&$8^{+1500}_{-8}\times10^{-3}$
&$5^{+95}_{-5}\times10^{-2}$\\
$g_{\omega^\prime_1\rho\pi}[\mbox{GeV}^{-1}]$&$-21^{+19}_{-32}\times
10^{-1}$&$<12$&$<50$&$-9^{+4}_{-2}$&$\sim-2$\\
$g_{\omega^\prime_1K^{\ast0}K^-\pi^+}$&$\sim0$&$-7^{+4}_{-50}\times10^2$
&$\sim-2000$&$\sim0$&$<500$\\
$g_{\omega^\prime_1\omega\pi^+\pi^-}$&$-120^{+60}_{-90}$&$-110\pm110$&
$\sim-70$&$\sim0$&$\sim-100$\\
$m_{\varphi^\prime_1}[\mbox{GeV}]$&$\sim1.5$&$\sim1.5$&$\sim1.5$&$\sim1.9$
&$\sim1.4$\\
$g_{\varphi^\prime_1K\bar K}$&$\equiv0$&$\equiv0$&$<15\times10^{-1}$
&$\equiv0$&$\equiv0$\\
$g_{\varphi^\prime_1K^{\ast0}K^-\pi^+}$&$\sim-300$&$\equiv-300$
&$\sim-400$&$<1200$&$\equiv0.$\\
\end{tabular}
\end{table}
\begin{table}
\caption{The same as in Table \protect\ref{tab1},
but for the  $\omega^\prime_2$ and
$\varphi^\prime_2$ resonances.\label{tab2}}
\begin{tabular}{cccccc}
final state&$\pi^+\pi^-\pi^0$&$\omega\pi^+\pi^-$&$K^+K^-$&$K^0_SK^\pm\pi^\mp$
&$K^{\ast0}K^-\pi^+$\\
\tableline
$m_{\omega^\prime_2}[\mbox{GeV}]$&$1.82^{+0.19}_{-0.15}$
&$1.84^{+0.10}_{-0.07}$&$1.78^{+0.17}_{-0.30}$&$\sim2.1$
&$1.88^{+0.60}_{-1.00}$\\
$\Gamma_{\omega^\prime_2e^+e^-}[\mbox{keV}]$&$10^{+9}_{-8}\times10^{-2}$
&$10^{+5}_{-3}\times10^{-2}$&$<6\times10^{-2}$
&$19^{+112}_{-19}\times10^{-2}$&$120^{+40}_{-50}\times10^{-2}$\\
$g_{\omega^\prime_2\rho\pi}[\mbox{GeV}^{-1}]$&$-7^{+4}_{-3}$
&$-7^{+3}_{-2}$&$<50$&$-11\pm2$&$-14\pm6$\\
$g_{\omega^\prime_2K^{\ast0}K^-\pi^+}$&$<3000$&$<1000$&$\sim0$
&$\sim0$&$31^{+40}_{-29}\times10^1$\\
$g_{\omega^\prime_1\omega\pi^+\pi^-}$&$90^{+100}_{-50}$&$90^{+40}_{-30}$&
$\sim0$&$\sim0$&$<2000$\\
Re$\Pi_{\omega^\prime_1\omega^\prime_2}[\mbox{GeV}]^2$&$<6\times10^{-1}$
&$<6\times10^{-1}$&$\sim0$&$\sim0$&$<20$\\
$m_{\varphi^\prime_2}[\mbox{GeV}]$&$2.8_{-1.1}$
\tablenote{$\chi^2$ is insensitive to greater values of the mass}&
$\sim1.9$&$\sim2.5$&
$2.6_{-0.3}$\tablenotemark[1]&$2.52^{+0.35}_{-0.27}$\\
$g_{\varphi^\prime_2K\bar K}$&$\equiv0$&$\equiv0$&$<3$
&$\equiv0$&$\equiv0$\\
$g_{\varphi^\prime_2K^{\ast0}K^-\pi^+}$&$\sim-100$&$\sim-20$
&$\sim-400$&$<600$&$-80^{+40}_{-60}$\\
Re$\Pi_{\varphi^\prime_1\varphi^\prime_2}[\mbox{GeV}]^2$&$\sim0.4$
&$\sim0$&$\sim0.1$&$<5$&$\sim0$\\
\end{tabular}
\end{table}
\narrowtext
\begin{table}
\caption{The values of $\chi^2$  per number of degrees
of freedom ($n_{\rm DOF}$) for each fitted channel.\label{tab3}}
\begin{tabular}{cc}
channel&$\chi^2/n_{\rm DOF}$\\
\tableline
$\pi^+\pi^-\pi^0$&40/36\\
$\omega\pi^+\pi^-$&6/5\\
$K^+K^-$&81/42\\
$K_SK^\pm\pi^\mp$&11/7\\
$K^{\ast 0}K^\pm\pi^\mp$&14/3\\
$K_LK_S$&5/7\\
\end{tabular}
\end{table}

\begin{references}
\bibitem{pdg}
R.~M.~Barnett {\it et al.} (Particle Data Group), Phys. Rev.
{\bf D54}, 1 (1996).
\bibitem{bityuk87}
S.~I.~Bityukov {\it et al.}, Phys. Lett. {\bf B188}, 383 (1987).
\bibitem{lipkin87}
F.~E.~Close and H.~J.~Lipkin, Phys. Rev. Lett. {\bf41}, 1263 (1978);
Phys. Lett. {\bf B196}, 245 (1987).
\bibitem{ach86}
N.~N.~Achasov, Pisma ZhETF {\bf43}, 410 (1986).
\bibitem{ach88}
N.~N.~Achasov and A.~A.~Kozhevnikov, Phys. Lett. {\bf B207}, 199 (1988);
{\it ibid} {\bf B209}, 373 (1988); Z. Phys. C{\bf 48}, 121 (1990).
\bibitem{kalash}
A.~Donnachie and Yu.~S.~Kalashnikova, Z. Phys. C{\bf59}, 621 (1993);\\
A.~Donnachie, Yu.~S.~Kalashnikova and A.~B.~Clegg, Z. Phys. C{\bf60},
187 (1993).
\bibitem{clegg94}
A.~B.~Clegg and A.~Donnachie, Z. Phys. C{\bf62}, 455 (1994).
\bibitem{page95}
F.~E.~Close and P.~R.~Page, Nucl. Phys. {\bf B443}, 233 (1995).
\bibitem{ach97a}
N.~N.~Achasov and A.~A.~Kozhevnikov,  Phys. Rev. D{\bf55}, 2663 (1997);
hep-ph/9609216.
\bibitem{nd}
S.~I.~Dolinsky {\it et al.}, Phys. Rep. C{\bf202}, 99 (1991).
\bibitem{dm2_3pi}
A.~Antonelli {\it et al.} Z. Phys. C{\bf56}, 15 (1992).
\bibitem{cmd}
L.~M.~Barkov {\it et al.} Pisma ZhETF {\bf46}, 132 (1987).
\bibitem{olya}
P.~M.~Ivanov {\it et al.} Phys. Lett {\bf 107B}, 297 (1981).
\bibitem{dm2_kk}
D.~Bisello {\it et al.} Z. Phys. C{\bf39}, 13 (1988).
\bibitem{dm2_kstk}
D.~Bisello {\it et al.} Z. Phys. C{\bf52}, 227 (1991).
\bibitem{dm2_kstkpi}
D.~Bisello {\it et al.} preprint LAL 90-35 (June 1990, unpublished).
\bibitem{godfrey85}
S.~Godfrey and N.~Isgur, Phys. Rev. D{\bf32}, 189 (1985).
\bibitem{ach84}
N.~N.~Achasov, S.~A.~Devyanin and G.~N.~Shestakov, Usp. Fiz. Nauk,
{\bf 142}, 361 (1984).
\bibitem{ach92}
N.~N.~Achasov {\it et al.}, Yadernaya Fizika, {\bf54}, 1097 (1991);
Int. Journ. Mod. Phys. {\bf A7}, 3187 (1992).
\bibitem{ach95}
N.~N.~Achasov and A.~A.~Kozhevnikov, Phys. Rev. D{\bf52}, 3119 (1995).
\bibitem{ach93}
N.~N.~Achasov and A.~A.~Kozhevnikov, Particle World {\bf3}, 125 (1993).
\bibitem{weiskopf}
R.~Van Royen and V.~F.~Weiskopf, Nuovo Cimento {\bf LA3}, 617 (1967).
\bibitem{novikov}
V.~A.~Novikov  {\it et al.}, Phys. Rep. C{\bf41}, 1 (1978).
\bibitem{odor}
K.~Fujikawa and P.~J.~O'Donnel, Phys. Rev. D{\bf8}, 3994 (1973):\\
S.~Matsuda, C.~Y.~Huang and S.~Oneda, Phys. Rev. D{\bf8}, 4133 (1973):\\
R.~Odorico, Phys. Rev. D{\bf 15}, 1384 (1977).
\bibitem{donnach87}
A.~Donnachie and H.~Mirzaie, Z. Phys. C{\bf33}, 407 (1987).
\bibitem{ivanov82}
P.~M.~Ivanov {\it et al.} Pisma ZhETF, {\bf36}, 91 (1982).
\bibitem{mane81}
F.~Mane {\it et al.} Phys. Lett. {\bf 99B}, 261 (1981).
\end{references}
\end{document}